\documentclass[11pt]{article}
\usepackage{amsfonts,amssymb,amsmath,epsfig,graphicx}
\usepackage{caption,subcaption,vmargin}
\setmarginsrb{2cm}{2cm}{2cm}{2.5cm}{0cm}{0cm}{0cm}{0.5cm}
\usepackage{cite}
\usepackage{tikz}
\usetikzlibrary{arrows}
\tikzstyle{block}=[draw opacity=0.7,line width=1.4cm]
\usetikzlibrary{positioning,arrows,chains,matrix,scopes,fit,decorations.markings,decorations}

\newtheorem{proposition}{Proposition}
\newtheorem{rmk}{Remark}
\newtheorem{lemma}{Lemma}

\newcommand{\bma}{\begin{pmatrix}}
\newcommand{\ema}{\end{pmatrix}}

\newcommand{\RR}{\mbox{${\mathbb R}$}}

\title{Vector NLS solitons interacting with a boundary\footnote{  Supported by NSFC (No. 11601312, 11631007, 11875040) and Shanghai Young Eastern Scholar   program (2016-2019).}}
\date{\empty}
\author{ Cheng Zhang\footnote{Corresponding author: {ch.zhang.maths@gmail.com}}\,,   \qquad Da-jun Zhang 
 \\     \sc \small Department of Mathematics, \small Shanghai University,   \small \sc Shanghai, 200444
}

\begin{document}
\maketitle
\begin{abstract}
We construct multi-soliton solutions of the $n$-component vector nonlinear Schr\"odinger equation on the half-line subject to two classes of integrable boundary conditions (BCs): the homogeneous Robin BCs and the mixed Neumann/Dirichlet BCs. The construction is based on the approach of dressing the integrable BCs: soliton solutions are generated in preserving the integrable BCs at each step of the Darboux-dressing process. Under the Robin BCs, examples, including boundary-bound solitons, are explicitly derived; under the mixed Neumann/Dirichlet BCs, the boundary can act as a polarizer that tunes different components of the vector solitons.  
  \vspace{.2cm}

\end{abstract}

\section{Introduction}
The concept of integrable boundary conditions (BCs), mainly developed by Sklyanin \cite{sklyanin1987boundary}, represents one of the most successful approaches to initial-boundary-value problems for two-dimensional integrable nonlinear PDEs. The idea lies on translating the integrability of soliton equations with boundaries into certain algebraic constraint known as reflection equation, cf.~\cite{Cherednik1, sklyanin1987boundary, ACC}. As consequences, classes of soliton models, restricted on a finite interval, are integrable subject to integrable BCs \cite{sklyanin1987boundary}. 

In this paper, we consider the focusing vector nonlinear Schr\"odinger equation (VNLS) equation, also known as the  Manakov model \cite{Mana}, restricted to the half-line space domain. The equation reads\begin{equation}
  \label{eq:Hnls}
  i {\mathbf r}_t + {\mathbf r}_{xx}+2\,  {\mathbf r}^\dagger \,H\, {\mathbf r}\,{\mathbf r}=   {\mathbf 0} \,,\quad {\mathbf r}:={\mathbf r}(x,t) \,,
\end{equation}
where ${\mathbf r} = (r_1,\cdots, r_n)^\intercal$,  $  {\mathbf 0}$ denotes the zero $n$-vector, and ${\mathbf r}^\dagger$ denotes the complex conjugate of ${\mathbf r}$. Each component $r_j$ is a complex field, and $H$ is an $n\times n$ positive definite Hermitian matrix modeling interactions among the components. There is a natural ${\cal U}(n)$-invariance of the model under the transformation $ {\mathbf r}\mapsto T\, {\mathbf r}$ where $T\in{\cal U}(n)$. Let $ T$ diagonalize $H$, then VNLS \eqref{eq:Hnls}, up to certain scaling, can be reduced to its standard form
\begin{equation}
  \label{eq:nls}
  i {\mathbf r}_t + {\mathbf r}_{xx}+2  {\mathbf r}^\dagger  {\mathbf r}\,{\mathbf r}=   {\mathbf 0} \,. 
\end{equation}

The VNLS equation is a vector generalization of the (scalar) NLS equation by allowing internal degrees of freedoms. Physically, it is a relevant model to describe optical solitons and collective states in low-temperature physics, cf.~\cite{SSS, CSCBE}; mathematically, the nontrivial interactions of vector solitons are related to  concepts of Yang-Baxter equation, cf.~\cite{Veselov, Tsuchida, APTN, CZNN}.   

Integrable BCs for VNLS, as well as soliton solutions to VNLS on the half-line, were derived in \cite{CZ} by means of a nonlinear mirror image technique \cite{biondini2009solitons} that extends the half-line space domain to the whole axis. However, there is a severe difficulty to construct $N$-soliton solutions on the half-line as the soliton data can only be computed in a recursive way. In practice, the computations are becoming increasingly complicated for $N\geq 2$ (see for instance  \cite[Conclusions]{CZ}).

We provide an efficient approach to deriving $N$-soliton solutions of VNLS on the half-line. The construction is based on the so-called {\em dressing the boundary}, introduced recently by one of the authors \cite{CZ2}. The essential steps are: given integrable BCs of VNLS (or any integrable PDEs), by generating soliton solutions using the Darboux-dressing transformations (DTs),  we look for those DTs that preserve the integrable BCs. This approach gives rise naturally to exact solutions of the underlying integrable model on the half-line. 
Compared to  the mirror image technique \cite{biondini2009solitons, CZ}, our construction has the advantages that i) it does not require extension of the space domain; ii) it leads efficiently to $N$-soliton solutions of VNLS on the half-line. 
Note that Fokas' unified transform method \cite{fokas1997unified, fokas2002integrable} represents a systematic approach to treating initial-boundary-value problems for integrable PDEs.  This method can be regarded as a generalization of the inverse transform transform (IST), cf.~\cite{kdv, ZS2, AKNS, faddeev}, and was already applied to NLS \cite{fokas2005nonlinear} and VNLS \cite{GLZ}. However, it is a difficult task to obtain exact solutions within the Fokas' method, although asymptotic solutions at large time could be derived. 

The outline and main results  of the paper are following. First, DTs for generating soliton solutions of VNLS are reviewed in Section $2$. Then we recall in Section $3$ results on integrable BCs for VNLS on the half-line \cite{CZ}. There are two classes of integrable BCs: the homogeneous Robin BCs and mixed Neumann/Dirichlet BCs.  In Section $4$, we apply the approach of dressing the boundary to VNLS on the half-line, which gives rise to explicit $N$-soliton solutions on the half-line. Our results provide clear  answer to the question of obtaining general $N$-soliton solutions in the presence
of a boundary \cite{CZ}. Moreover, we can construct stationary vector solitons subject to the Robin BCs at the boundary. These correspond to boundary-bound solitons.     In Section $5$, we provide explicit examples of vector solitons interacting with the boundary. In particular, by combining the effects of the mixed Neumann/Dirichlet BCs and the ${\cal U}(n)$-invariance of VNLS, the boundary can act as a polarizer that tunes components of solitons after interacting with the boundary. 
\section{DTs and soliton solutions}
 The $n$-component VNLS equation is equivalent to the compatibility of the linear differential system
 \begin{equation}
   \label{eq:oLax}
  \Phi_x = U\, \Phi\,,\quad   \Phi_x = V\, \Phi\,.
\end{equation}
Here, $U, V$, commonly known as Lax pair, are $(n+1\times n+1)$ matrix-valued functions
\begin{equation}
U=  -i \lambda \Sigma + Q\,,\quad  V= -2i \lambda^2 \Sigma + 2\lambda Q-i Q_x\,\Sigma -i Q^2\,\Sigma\,,
\end{equation}
where $\lambda$ is the spectral parameter, and $\Sigma$ and $Q$ are block matrices
  \begin{equation}
    \label{eq:1}
  \Sigma = \bma I_n &  {\mathbf 0} \\ {\mathbf 0}^\intercal & -1 \ema  \,,\quad Q = \bma  O_n & {\mathbf r} \\  -{\mathbf r}^\dagger & 0\ema\,,
  \end{equation}
with $I_n, O_n$ being the identity and zero square matrices of size $n$ respectively. There is a natural {gauge} group acting on the  Lax pair~\eqref{eq:oLax}
\begin{equation}
  \label{eq:gauge}
  \widetilde{U} = G\, U \,G^{-1}+ G_x\,G^{-1}\,,\quad \widetilde{V} = G\, V \,G^{-1}+G_t\,G^{-1}\,, 
\end{equation}
and DTs can be represented by $G$ that preserves the forms of $U,V$ by extracting the pole structures, cf.~\cite{ZS4, matveev1991darboux, babelon, CJL}.   
A one-step DT for VNLS amounts to the map $\Phi\mapsto \widetilde{\Phi} =D[1] \Phi $, where $D[1]$ is called dressing factor of degree $1$ 
\begin{equation}
  \label{eq:Dr}
  D[1](\lambda) = I_{n+1} + ( \frac{\lambda^*_1-\lambda_1}{\lambda-\lambda_1^*})\Pi_1\,,\quad \Pi_1 = \frac{\Psi_1 \Psi_1^\dagger}{\Psi_1^\dagger \Psi_1}\,.
\end{equation}
Here $\Psi_1$ is a particular solution of the {\em undressed} Lax pair \eqref{eq:oLax} associated to $\lambda_1$.  Having a set of $N$ particular solutions $\{\Psi_j,\lambda_j\}$, $j=1,\dots, N$, one can  iterate the DTs and construct the dressing factor $D[N]$ of degree $N$. For simplicity $\Psi_j$'s are assumed to be vectors (of rank $1$). In the IST formalism, $D[N]$ plays the role of the scattering matrix: one adds a pair of complex zero/pole $\{\lambda_j,\lambda_j^*\}$ to the scattering system at each step of the DTs
. 
There are two important properties of DTs:  1) the {\em Bianchi permutativity} meaning that the order of adding $\Psi_j$ is irrelevant; and 2) the action of $D[N](\lambda)$ can be expressed in compact forms (usually in terms of determinant structures).

Since we are focusing on soliton solutions, the zero seed solution    ${\mathbf r} =\mathbf 0 $ is imposed in the undressed Lax pair. Without loss of generality, let  $\Psi_j$'s be  in the forms
\begin{equation}
  \label{eq:psij}
  \Psi_j = e^{-i(\lambda_j x+2\lambda_j^2 t) \Sigma}\bma {\mathbf b}_j \\ 1 \ema \,,  
\end{equation}
where $ {\mathbf b}_j$'s are constant complex $n$-vectors called {\em norming vectors}. Encoding now the soliton data into $\{{\mathbf b}_j,\lambda_j\}$, $j=1,\dots, N$, with $\lambda_j$'s being distinct, the $N$-soliton solutions are, cf.~\cite{faddeev, babelon} 
\begin{equation}
  \label{eq:sol22}
  r_\ell= \frac{2i }{\det M}\begin{vmatrix} 0& \begin{matrix}   1 & \cdots & 1  \end{matrix}  \\ \begin{matrix}  \beta_{1;\ell} \\ \vdots \\ \beta_{N;\ell} \end{matrix} & M       \end{vmatrix} \,, \quad \ell = 1,\dots, n\,,
\end{equation}
 where $r_\ell$ is the $\ell$-th component of ${\mathbf r}$, and $\beta_{j}:= \beta_{j}(x,t)  =  e^{-2i(\lambda_j x+2\lambda_j^2 t) } {\mathbf b}_{j} $ with $ {\beta}_{j;\ell}$ being its $\ell$-th component. The $N\times N$ matrix $M$ has components 
 $M_{j\ell}=\frac{\beta^\dagger_\ell\beta_j+1}{\lambda_j^*-\lambda_\ell}$.
 As an illustration, the one-soliton data  $\{{\mathbf b}_1,\lambda_1\}$, with  $\lambda_1 = \frac{1}{2}(\mu_1+i\nu_1)$, $\nu_1>0$, lead to one vector soliton solution
 \begin{equation}
   \label{eq:1ss}
{\mathbf r}(x,t)=\mathbf{p}\, \frac{\nu_1 e^{-i(\mu_1 x+(\mu_1^2-\nu_1^2)t-\pi)}}{\cosh(\nu_1 (x+2\mu_1 t+\Delta x))}  \equiv \mathbf{p} \,r(x,t)\,.
 \end{equation}
Here $ \Delta  x =\frac{\log |\mathbf  {b}_1|}{\nu_1}$, $  {\mathbf p} =\frac{\mathbf {b}_1}{|\mathbf {b}_1|}$, the solution ${\mathbf r}(x,t)$ is composed by the usual scalar NLS soliton solution $r(x,t)$, where the amplitude  $\nu_1$ and the velocity  $2\mu_1$  are controlled by the imaginary and real parts of $\lambda_1$ respectively,  and a unit  {\em polarization}  vector $\mathbf p$. 
 \section{Integrable BCs for VNLS}
Now we restrict the space domain of VNLS to the positive semi-axis. Integrable BCs for the half-line  VNLS was investigated in \cite{CZ} (see also \cite{habibullin1995integrable} in which only the vector Robin BCs were derived). The integrability in the presence of a boundary was translated into a constraint on the $t$-part of the Lax pair
\begin{equation}
  \label{eq:bccs}
K(\lambda)V(- \lambda)\rvert_{x=0} =V(\lambda)\rvert_{x=0} \,K(\lambda)\,.
\end{equation}
Here the  {\em boundary matrix}  $K(\lambda)$ is assumed to be nondegenerate. 
As solutions of the boundary constraint \eqref{eq:bccs},  two classes of BCs were obtained \cite{CZ}: i) the  homogeneous vector Robin BCs:
\begin{equation}
  \label{eq:vrbcs}
    ({\mathbf r}_x - 2\alpha {\mathbf r})\rvert_{x=0}={\mathbf 0}\,, \quad \alpha \in \RR\,,\end{equation}
  having the boundary matrix
  \begin{equation}
    \label{eq:faaaa1}
    K(\lambda)  = \bma f_a(\lambda)I_n &  {\mathbf 0} \\ {\mathbf 0}^\intercal & 1 \ema  \,,\quad f_\alpha(\lambda) =\frac{i\alpha+\lambda }{i\alpha-\lambda}\,.
  \end{equation}
 The real parameter $\alpha$ controls the boundary behavior: the Neumann (${\mathbf r}_x\rvert_{x=0} = 0)$ and Dirichlet BCs  (${\mathbf r}\rvert_{x=0} = 0$) appear as special cases of \eqref{eq:vrbcs} as $\alpha = 0$ and $|\alpha| \to \infty$ respectively;  ii) the mixed Neumann/Dirichlet (mND) BCs:
  \begin{equation}
    \label{eq:KmND}    r_{\ell_x}\rvert_{x=0} = 0\,,\quad r_{j}\rvert_{x=0} = 0\,, \quad \text{for }j\neq\ell,
  \end{equation}
where $r_\ell$ is the $\ell$-th component of $\mathbf r$. Accordingly, one has the boundary matrix
  \begin{equation}
    \label{eq:bmax1}
    K =  \text{diag}(\delta_1, \dots, \delta_n, 1)\,,\quad  \delta_\ell =1 , \quad \delta_j =-1,   \quad \text{for }j\neq\ell, 
  \end{equation}
  where $+/-$ sign of $\delta_\ell$ corresponds to Neumann/Dirichlet BCs.
\begin{rmk}
The boundary constraint  \eqref{eq:bccs} was derived in \cite{CZ} by considering the space-reverse symmetry of VNLS as a B\"acklund transformation. The same constraint was also introduced in Fokas' unified transform known as {\em linearizable BCs}. Note that the boundary matrix $K$ is related to a far-reaching context as it represents solutions of the classical and quantum reflection equations \cite{sklyanin1987boundary, Cherednik1, ACC}. 
\end{rmk} 
\begin{rmk}
\label{rm:r2}The integrable BCs are compatible with the ${\cal U}(n)$-invariance of the VNLS equation. The transformation ${\mathbf r}\mapsto \widetilde{{\mathbf r}} =T\, {\mathbf r} $, $T\in {\cal U}(n)$, is trivial to the Robin BCs, because a collective change the components of ${\mathbf r}$ takes place at the boundary under $T$. However, $T$ induces a nontrivial effect under the mND BCs: since the components of ${\mathbf r}$ can interact differently with the boundary in two ways that are Neumann and  Dirichlet BCs, the action of $T$ can mix the two interactions  and make transmissions among the different components appear.  This transmission phenomena have the interpretation that the boundary acts as a  ``polarizer" tuning the polarizations of the incoming solitons, after interacting with boundary,  changes of the polarizations among the solitons take place \cite{CZ}. \end{rmk}
\section{Dressing the boundary}
The integrable BCs for the VNLS equation on the half-line are completely determined by the $t$-part of the Lax pair through the boundary constraint \eqref{eq:bccs}. By {\em dressing the boundary} \cite{CZ2}, we mean that in the process of DTs to generate exact solutions, the boundary constraint is preserved at each step of the DTs. By construction, this leads to exact solutions of VNLS subject to the integrable BCs. In practice this requires to find out appropriate particular solutions in DTs.
\begin{lemma}[Dressing the boundary]\label{lem:1}
Let  $U,V$ be the undressed Lax pair. Assume that $V$ satisfies the boundary constraint \eqref{eq:bccs}, and that the Lax pair admits a pair of particular solutions  $\Psi_j, \widetilde{\Psi}_j $, associated to  $\lambda_j, \widetilde{\lambda}_j$ respectively (assume $\lambda_j$ is not pure imaginary), such that 
\begin{equation}\label{eq:ccbcs11}
\widetilde{\Psi}_j\vert_{x=0} = K(-\lambda_j ) \Psi_j\vert_{x=0}\,,\quad \widetilde{\lambda}_j = - \lambda_j\,,  
\end{equation}
where $K(\lambda)$ is the boundary matrix, then the boundary constraint  \eqref{eq:bccs} is preserved after dressing $V$ using  $\Psi_j, \widetilde{\Psi}_j $. 
\end{lemma}
 The proof is closely related to the structure of dressing factors. Similar statements can be found in \cite{CZ2} for the scalar case. In order to obtain exact solutions on the half-line, it remains to find the paired particular solutions   $\Psi_j, \widetilde{\Psi}_j $ satisfying \eqref{eq:ccbcs11}. 

\begin{proposition}[$N$-soliton solutions on the half-line]
  \label{prop:1}
Let  $\{{\mathbf b}_j,\lambda_j\} $ and  $\{\widetilde{{\mathbf b}}_j,\widetilde{\lambda}_j\} $, $j=1,\dots,N$, be two sets of $N$-soliton data. Assume that $\widetilde{\lambda}_j = -\lambda_j$ ($\lambda_j$ is not pure imaginary)  and  $\widetilde{\mathbf b}_j  = B(-\lambda_j) {\mathbf b}_j$ with $B(\lambda) = f_\alpha(\lambda)I_n$ ($f_\alpha(\lambda)$ defined in \eqref{eq:faaaa1}), then the so-construct  solutions restricted to $x\geq 0$ correspond to $N$-soliton solutions on the half-line subject to the Robin BCs \eqref{eq:vrbcs}; if $B=-\text{diag}(\delta_1, \dots, \delta_n)$, $  \delta_\ell =1$, $\delta_j =-1$, for $j\neq\ell $, then the solutions restricted to $x\geq 0$ satisfy   the mND BCs \eqref{eq:KmND}. 
\end{proposition}

The proof is a direct consequence of Lemma~\ref{lem:1} by taking into account the forms of the particular solutions \eqref{eq:psij}. 
Dressing the Lax pair using the $N$-paired soliton data  $\{{\mathbf b}_j,\lambda_j\} $ and  $\{\widetilde{{\mathbf b}}_j,\widetilde{\lambda}_j\} $ gives rise to $2N$-soliton solutions on the whole line, and the requirements that $\widetilde{\lambda}_j = -\lambda_j$ create solitons with opposite velocities. By restricting the space-domain to the positive semi-axis, the BCs appear as interactions of solitons with opposite velocities at $x=0$, then one obtains $N$-soliton solutions on the half-line. Although this whole-line picture helps to interpret interactions of solitons as BCs,  the derivation of soliton solutions on the half-line can be restricted to $x\geq 0$. This is in contrast to the nonlinear mirror-image technique \cite{biondini2009solitons, CZ} where an extended potential to the whole-line is required.

Note that in the above construction, pure imaginary  $\lambda_j$'s, corresponding to stationary solitons are excluded. By dressing the boundary, we can also construct stationary solitons satisfying the Robin BCs. These are boundary-bound solitons on the half-line. 
\begin{proposition}[Boundary-bound solitons]
  \label{prop:2}
Let ${\mathbf b}$ be any unit complex $n$-vector and $\{\widetilde{{\mathbf b}}_j,\lambda_j\} $,   $j=1,\dots,N$, be a set of $N$-soliton data. Assume that $\lambda_j$'s are pure imaginary numbers and distinct,  and for given $\alpha$, satisfy $f_\alpha(\lambda_j)<0$ ($f_\alpha(\lambda)$ defined in \eqref{eq:faaaa1}). Moreover, assume the following forms of the norming constants  $\widetilde{{\mathbf b}}_j= {\mathbf b} \left(\sqrt{-f_\alpha(\lambda_j)}\right)^{(-1)^N}$, then the so-constructed solutions restricted to $x\geq 0$ correspond to $N$-stationary solitons on the half-line subject to the Robin BCs \eqref{eq:vrbcs}. 
\end{proposition}

Again the restriction on the soliton data follows the idea of dressing the boundary: the  boundary constraint \eqref{eq:bccs} is preserved  at each step of the DTs. In computing the boundary-bound solitons, the expressions for the norming constants are different for the odd and even soliton numbers. One also excludes the situation where the stationary solitons are subject to the Dirichlet BCs by assuming  $f_\alpha(\lambda_j)<0$. Note that for the  scalar NLS case, the boundary-bound states were investigated in  \cite{bion2, CZ2}. One can put the stationary and moving solitons  together  by combining the associated soliton data. Due to the Bianchi  permutativity of DTs, the order of adding the soliton data is irrelevant.   

\section{Examples of VNLS soliton interacting with a boundary}
\begin{figure}[h!]
  \centering
  \includegraphics[width=0.35\linewidth]{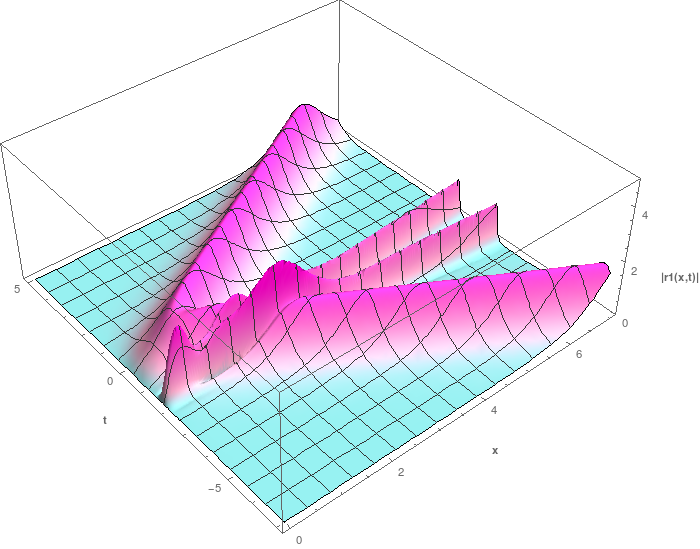} \hspace{-.1cm}
  \includegraphics[width=0.35\linewidth]{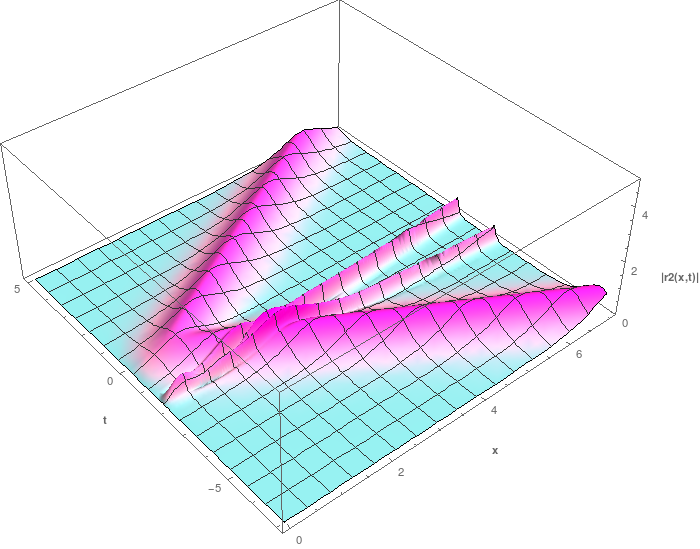}
  \caption{Two vector solitons vanish at the boundary subjec to the Dirichlet BCs}
  \label{fig:1}
\end{figure}

\begin{figure}[h!]
  \centering
  \includegraphics[width=0.35\linewidth]{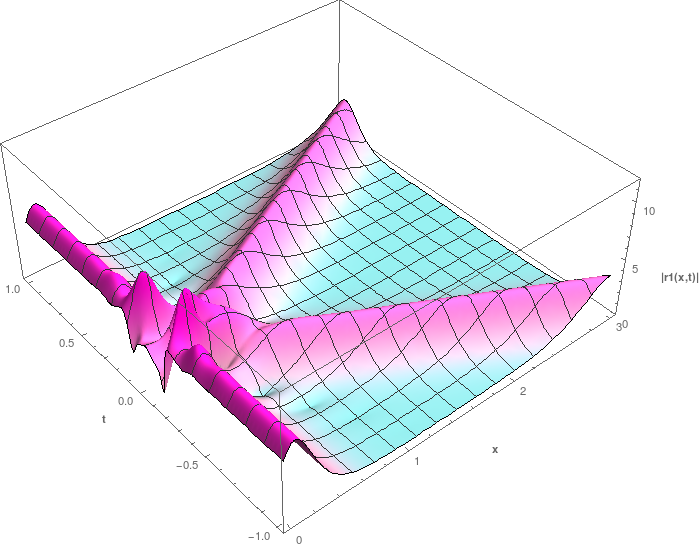} \hspace{-.1cm}
  \includegraphics[width=0.35\linewidth]{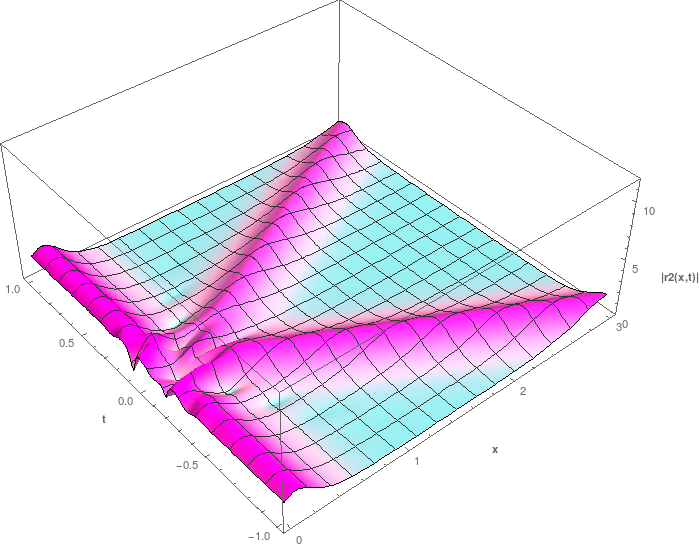}
  \caption{One soliton interact with a boundary-bound soliton subject to the Robin BCs ($\alpha=2$)}
  \label{fig:2}
\end{figure}

\begin{figure}[h!]
  \centering
  \includegraphics[width=0.35\linewidth]{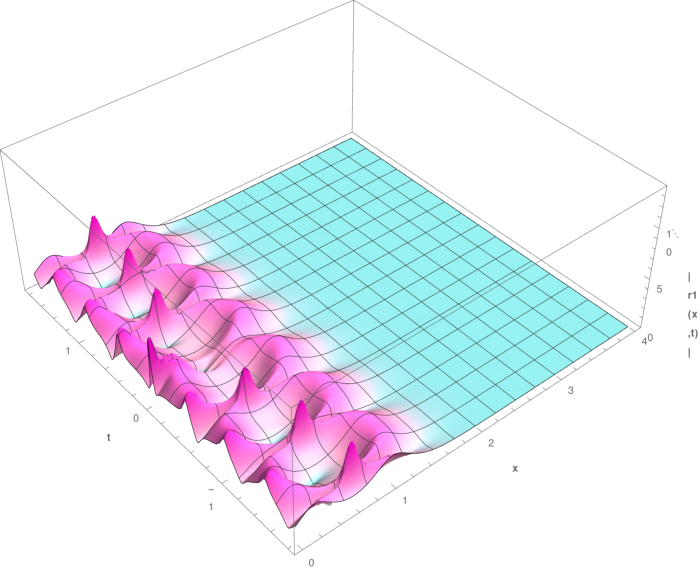} \hspace{-.1cm}
  \includegraphics[width=0.35\linewidth]{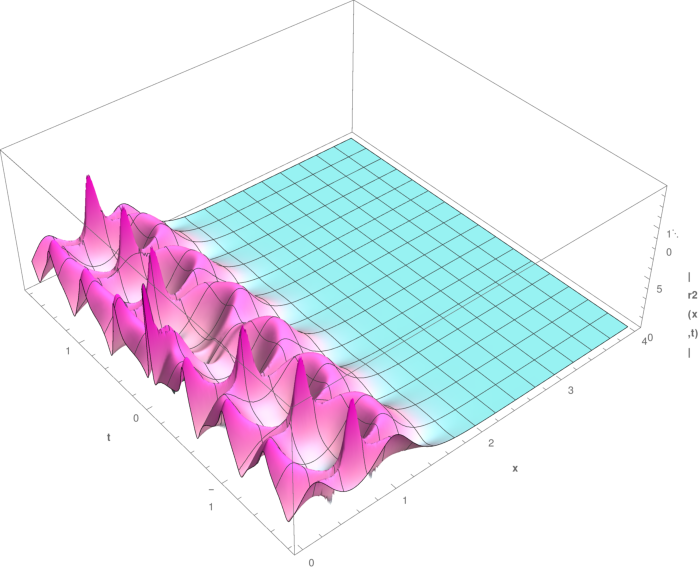}
  \caption{Three stationary solitons interfere with themselves at the boundary subject to the Robin BCs ($\alpha=2$)}
  \label{fig:3}
\end{figure}

It is straightforward to apply Prop.~\ref{prop:1} and~\ref{prop:2} to obtain soliton solutions of VNLS on the half-line. Fix $n=2$, three examples under the Robin BCs are shown in  Fig.~\ref{fig:1}-\ref{fig:3}. The left and right figures represent respectively the norms of the $1$st and $2$nd components of the solutions. 

\begin{figure}[h!]
  \centering
  \includegraphics[width=0.35\linewidth]{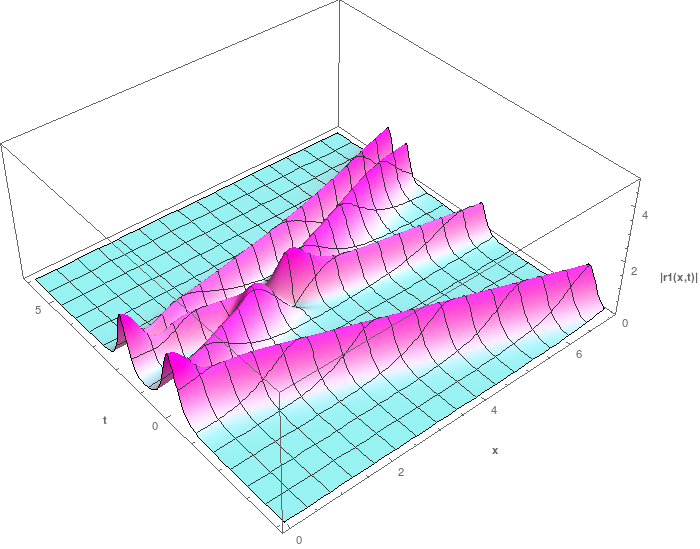} \hspace{-.1cm}
  \includegraphics[width=0.35\linewidth]{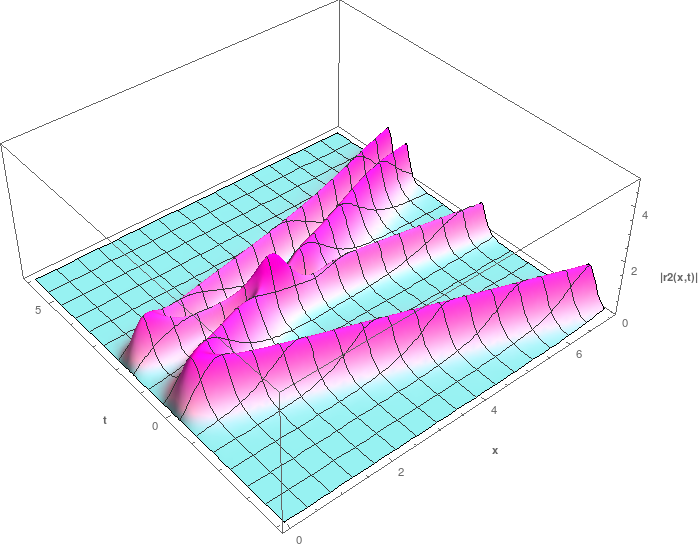}
  \caption{Two solitons interact with the mixed Neumann ($1$st component) and Dirichlet ($2$nd component) BCs}
  \label{fig:4}
\end{figure}

\begin{figure}[h!]
  \centering
  \includegraphics[width=0.35\linewidth]{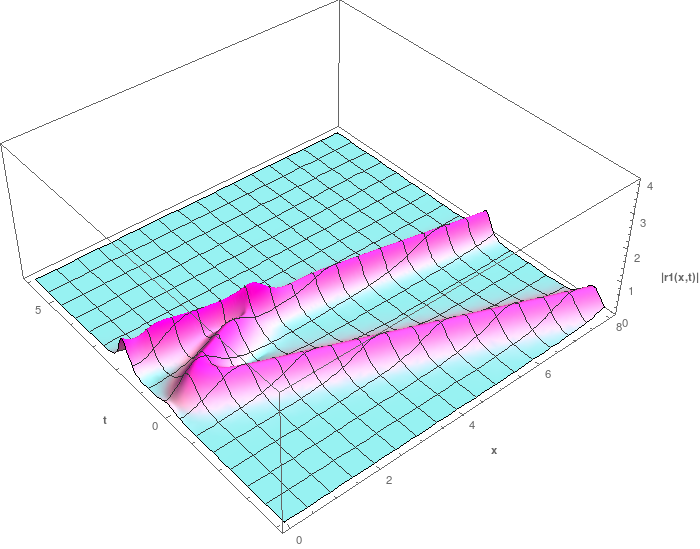} \hspace{-.1cm}
  \includegraphics[width=0.35\linewidth]{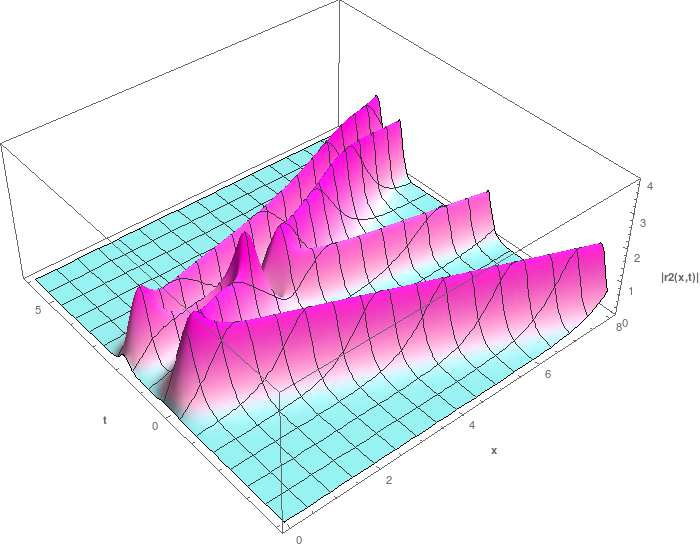}
  \caption{Polarizer effect: the boundary tunes the polarizations, and the $1$st component becomes vanishingly small after interacting with the boundary }
  \label{fig:5}
\end{figure}

As to the mND BCs, fix $n=2$ , and let the transformation matrix $T\in{\cal SU}(2)$ (following Remark~\ref{rm:r2}) parameterized by three parameters $\omega, \theta, \xi$ be in the form
\begin{equation}T= \bma e^{i\omega}\cos \theta &  -e^{i\xi}\sin \theta \\ e^{-i\xi}\sin \theta  &  e^{-i\omega}\cos\ema \,. \end{equation}
Clearly,  ${\mathbf r}\mapsto T{\mathbf r}$ induces a mixture of components of ${\mathbf r}$ at the boundary. In the computations of half-line soliton solution, this amounts to    $B\mapsto TBT^{-1}$ for $B$  defined in Prop.~\ref{prop:1}.   Examples of two solitons interacting with a mND boundary is shown below.  Having $B = \text{diag}(1,-1)$ gives rise to solitons with the $1$st component subject to Neumann BCs and $2$nd to Dirichlet BCs (see Fig.~\ref{fig:4}); under the action of $T$, for certain choices of the parameters, one can make one component of the outgoing solitons vanishingly small\footnote{The complete analysis, requiring some asymptotic estimations of the solutions as $t\to \pm \infty$, is omitted here} (see Fig.~\ref{fig:5}). In other words, the boundary polarizer switches off the $1$st component after solitons interacting with the boundary. 

\end{document}